\let\accentvec\vec
\documentclass[iop]{emulateapj}

\let\vec\accentvec
\usepackage{amsmath}
\usepackage{amssymb}
\usepackage{graphicx}
\usepackage{hyperref}
\usepackage{tocbibind}

\begin{document}
\newcommand{\beq}{\begin{equation}}
\newcommand{\eeq}{\end{equation}}
\newcommand{\ben}{\begin{enumerate}}
\newcommand{\een}{\end{enumerate}}
\newcommand{\bit}{\begin{itemize}}
\newcommand{\eit}{\end{itemize}}
\newcommand{\barr}{\begin{array}}
\newcommand{\earr}{\end{array}}
\newcommand{\D}{\partial}
\newcommand{\DD}{\frac}
\newcommand{\mm }{\mathrm}
\newcommand{\DroII}{\overline{\overline{\rm D}}}
\newcommand{\DroI}{{\overline{\rm D}}}
\long\def\symbolfootnote[#1]#2{\begingroup%
\def\thefootnote{\fnsymbol{footnote}}\footnote[#1]{#2}\endgroup}

\title{Advection schemes for capturing relativistic shocks with high Lorenz factors}

\author{Hujeirat, A.A.\altaffilmark{1,2}  and Fehlmann, S.\altaffilmark{2}}

\altaffiltext{1}{ IWR - Interdisciplinary Center for Scientific Computing, Heidelberg University, INF 368, 69120 Heidelberg, Germany}
\altaffiltext{2}{Departement Physik, University of Basel, Switzerland  }

\begin{abstract}
  {{ Jet-plasmas emanating from the vicinity of relativistic objects and in gamma-ray bursts have been observed
  to propagate with Lorentz factors  laying  in the range between one and several hundreds.
  On the other hand, the numerical studies of such flows have been focussed so far mainly on the
  lowest possible range of Lorentz factors $\Gamma,$ specifically, on the regime $1\leq \Gamma \leq 5.$  Therefore, relativistic flows with
  high $\Gamma-$factors have poorly studied, as most numerical methods are found to encounter
  severe numerical difficulties or even become numerically unstable for $\Gamma \gg 1.$}

    In this paper we present an implicit numerical advection scheme for modeling the propagation
    of relativistic plasmas with shocks,  discuss its consistency with respect to both the internal and total
  energy formulation in general relativity.
  Using the total energy formulation, the scheme is found to be viable for modeling
   moving shocks with moderate Lorentz factors, though with relatively small Courant
   numbers.
   In the limit of high Lorentz factors, the internal energy formulation in combination
   with a fine-tuned artificial viscosity is much more robust and efficient.
   We confirm our conclusions by performing test calculations and compare the results
   with analytical solutions of the relativistic shock tube problem.
   The aim of the present modification is to enhance the robustness of the general relativistic implicit
   radiative MHD solver:  GR-I-RMHD (http://www1.iwr.uni-heidelberg.de/groups/compastro/home/gr-i-mhd-solver)
   and extend  its range of applications into the high $\Gamma-$regime.}
\end{abstract}

\keywords{Relativity: numerical,  --- Hydrodynamics: relativistic, advection schemes
 --- Numerical methods: shock capturing, numerical accuracy  --- fluids: Euler and Navier-Stokes equations}

\section{Introduction}
Plasmas moving at relativistic speeds have been observed in diverse
astrophysical phenomena, such as in supernova explosions, in jets emanating
from around neutron stars, in microquasars and in active galactic nuclei
\citep[see][and the references therein]{Livio2004, Marscher06, Fender2007}.
The bulk Lorentz factor, $\Gamma,$ of the observed jet-plasmas, in most cases,
 has been found to
increase with the mass of the central relativistic object.
Jets ejected by neutron stars generally propagate with $\Gamma_{NS}$
between 1 and 3,  in   microquasars $~\Gamma_{\mu QSO}$ is between 1 and 4 whereas
in AGNs/QSOs $~\Gamma_{AGN}$ is between 5 and 10.
Gamma-ray bursts are considered to be an extreme case, as the Lorentz factors here
is considered to be of the order of several hundreds \citep[e.g.,][]{Piran2005}.

Jet-plasmas considered to decelerate via self-interaction or  interaction with
the surrounding media in the form of shocks and eventually
become efficient sources for the production of the observed high energy gamma-rays
and probably for the energetic cosmic rays \citep{Sikora1994}.

Modeling the formation of relativistic shock by means of HD and MHD simulation
has been the focus of  numerous studies during the the last two decades
\citep[][see also the references therein]{Hawley_etal1984b, Aloy1999, Font_etal2000, Hujeirat2003,MartiMueller2003, DeVilliersHawley2003,
Gammie_etal2003, Anninos, Komissarov2004, DelZanna2007, Mignone_etal2007, Hujeirat2008}.

These studies have enriched the field of computational astrophysics
with numerical techniques and useful strategies for accurate capturing of
relativistic shock fronts
using both the internal and total energy formulation \citep[e.g.,][]{DeVilliersHawley2003, Mignone_etal2007}.
{Nevertheless, most of these methods  encounter numerical difficulties
when  modeling relativistic flows with high Lorentz factors, i.e., $~\Gamma \geq 5.$
In the literature, results of relativistic shock tube problems with merely
moderate $~\Gamma-$foctors have been presented \citep[see][and the references therein]{Hawleyetal1984, Aloy1999, Zhang2006, Mignone_etal2007}.

{ The numerical difficulties associated with modeling flows with high Lorentz factors originate
from the small initial velocity errors in the momentum equations.
These errors correlate with $~\Gamma^2,$ hence they are highly non-linear and may be strongly magnified
as the time-iteration proceeds. This may give rise to a significant over/under estimation of the
other variables.
  Recipes, such as  further reducing the grid spacing and/or the time-step size may lead to a stagnation
  of the solution procedure, in particular if the  method used is conditionally stable.}

{
On the other hand, only a limited effort has been made to develop time-implicit numerical solvers for
simulating relativistic flows in astrophysics.
 The main reason therefor is that most astrophysical flows are intrinsically time-dependent with strong spatial
 variations, which implies
 that the numerical scheme must be highly accurate in time and space. However, the leading temporal errors
         generally scale as the time step size $\sim\delta t$ or the power of it. Thus $\delta t$ must be sufficiently small
         and in most cases,  it must be even smaller than the sound-crossing time between the boundaries of a single
         finite volume cell.
         This is equivalent to require that the Currant number be smaller than unity and therefore ending up with the
         stability requirement of conditionally stable methods. In this regime explicit methods are much more efficient than
         their implicit counterparts, as implicit methods require the inversion of a matrix $A_{imp}$ (e.g. Eq. \ref{Matrix_Impl})
         in order to solve the set of linear equations $A x = b$ that corresponds to
          the linearized hydrodynamical equations in the finite space.
         \beq
         A_{imp} \sim \left(
         \begin{matrix}
           \DD{a}{\epsilon} & \epsilon & \epsilon & \epsilon \\
           \epsilon & \DD{a}{\epsilon} & \epsilon & \epsilon \\
             &  & \ddots &  \\
           \epsilon & \epsilon & \epsilon & \DD{a}{\epsilon}
         \end{matrix}\right)
         \label{Matrix_Impl}
         \eeq
 On the other hand, it can be easily verified that one may obtain almost the same results when $A_{imp}$ is replaced by $A_{exp}=\DD{a}{\epsilon}~ I$,
 where I is the identity
 matrix\footnote{The similarity between the real Jacobian $\mathcal{J}$ and the matrices
         $A_{imp}$ and $A_{exp}$  applies for a sufficiently small $\delta t$.}.
   $A_{exp}$ is the pre-conditioner generally used by explicit methods for solving the same set of
 equations, thereby avoiding the formidable arithmetic operations associated with the inversion procedure of $A_{imp}.$

 However, most astrophysical flows are by nature dissipative, radiative and magnetic-diffusive and therefore are well-described by the
 the generalized Navier-Stokes (-NS) rather than the Euler equations.
 Such physically non-ideal effects generally do not operate uniformly in the domain of calculations and they may occur also on longer or shorter time scales than the dynamical one. As a consequence, the NS-solvers should not only be unconditionally stable, but they must be sufficiently robust to treat
 limiting cases such as Euler type-flows, event hough their efficiency for these specific cases might be too low compared to explicit methods.}

, hence the purpose of the present paper. Moreover,
we show that our time-implicit method in combination with a modified third order MUSCL
advection scheme is capable of modeling shocks propagating with Lorentz factors
 $~\Gamma \geq 10$ with time step sizes corresponding to Courant numbers larger than
 100.

In Sec. 2 we describe the hydrodynamical equations solved in the present study.
The solution method relies on using the 3D axi-symmetric general relativistic implicit
radiative MHD solver (henceforth GR-I-RMHD), which has been described in details in
a series of articles \citep[e.g., ][]{Hujeirat2008, Hujeirat2009}.
In Sec. (3) we just focus on several new aspects of the advection scheme. The results
of the verification tests are presented in Sec. (4) and end up with Sec. (5), where
the results of the present study are summarized.
\begin{figure*}
\centering {\hspace*{-0.2cm}
\includegraphics*[width=8.5cm] {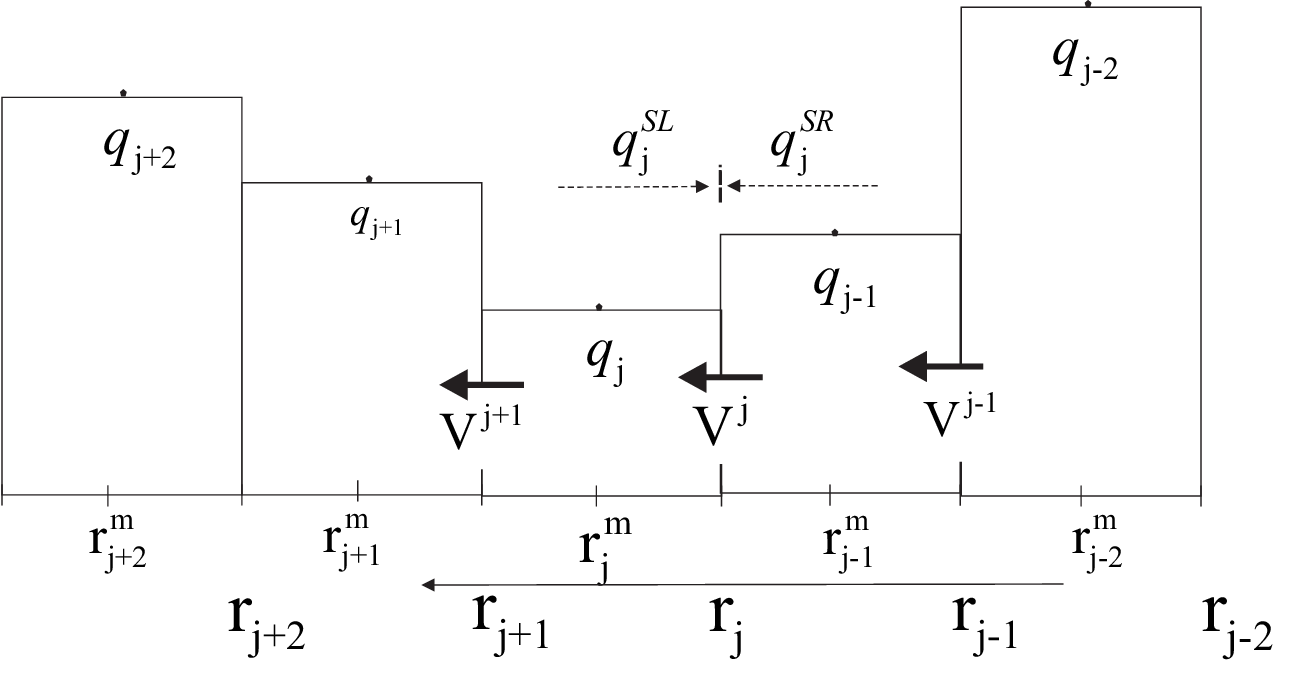}}
\caption{\small  A schematic description of several finite volume cells, their boundaries
 and the location of the scalar quantities "q" and the velocities.} \label{Fig1}
\end{figure*}

\section{The 1D general relativistic hydrodynamical equations}

In the present study we consider the set of Euler equations in one-dimension and in
flat spacetime:
\beq
\barr{l}
 \rm {\D_t D} + \nabla_r \cdot (D\, V^r) = 0 \\
 {\D_t M_r} + \nabla_r \cdot (M_r V^r)= - {\D_r P} \\
{\D_t \mathcal{E}^\mm{tot} }+ \nabla_r \cdot ([\mathcal{E}^\mm{tot} + P] V^r) = 0,
\earr
\label{Eq1}
\eeq
{where $\rm \nabla_r \cdot (\doteq \DD{1}{\sqrt{-g}} {\D}_r \sqrt{-g}),~~$ is the radial component of the diveregence operator in flat space with the signature (-,+,+,+) and ,,g`` is the determinant of the coefficient matrix corresponding to the metric.
$\rm D (\doteq \rho u^t)$ is the relativistic density,
$\rm{M}_r$ is the radial component of the 4-momentum: \(\rm M_\mu \doteq {\DroI}u_\mu,\) where
$u_\mu$ is the 4-velocity,
  $\rm u^t$ is the time-like
velocity\footnote{$\rm u^t$ is also the general relativistic Lorentz factor, which
reduces to the classical $\Gamma$ in flat space.}, \(\rm  V^r (= u^r/u^t)\) is
the transport velocity, $\rm {\DroI} \doteq D h,$ and  ``h" is the relativistic enthalpy $\rm h=c^2 + \varepsilon + P/\rho$.
$\rm \mathcal{E}^\mm{tot} = (u^t)^2 \rho h -P$  is the total energy which is the sum of kinetic and
 thermal energies
of the gas. ``P" is the pressure of ideal gas:
$\rm P = (\gamma-1)\rho \varepsilon,$
$\varepsilon$ and $\gamma$ denote the internal energy and the corresponding adiabatic index, respectively. \\
In the rest of this paper, we will set c=1.}

 The reader is referred to  Sec. (2) of Hujeirat et al. (2008), where the general relativistic
hydrodynamical equations and their derivations in the Boyer-Lindquist coordinates
are described. However, we continue to use these coordinates, though in the limit
of  flat space.

In the case that the mechanical energy is conserved, then the total energy reduces
to an evolutionary equation for the internal energy:

\beq
\rm
{\D_t {\cal{E}}^d} + \nabla_r \cdot({\cal{E}}^d V^r)=
-(\gamma-1)\,\DD{{\cal{E}}^d}{u^t} [\D_t u^t + \nabla_r \cdot (u^t~V^r)]
               \label{Eq5in}
\eeq
where $\rm {\cal{E}}^d = u^t P/(\gamma-1).$

Furthermore, there is a remarkable similarity of how Eq. (\ref{Eq5in}) deals
with $\rm {\cal{E}}^d$ and $u^t.$ Taking this similarity into account,
the equation can be re-formulated to have the following form:
\beq
\rm
{\D_t {\bar{\cal{E}}}^d} + \nabla_r \cdot ({\bar{\cal{E}}}^d V^r)=
-\gamma D (\nabla_r \cdot V^r),
               \label{Eq5in2}
\eeq
where $\rm \bar{\cal{E}}^d = {D\log}[{\cal{E}}^d (u^t)^{\gamma-1}].$

\begin{figure*}[htb]
\centering {\hspace*{-0.2cm}
\includegraphics*[width=10.5cm] {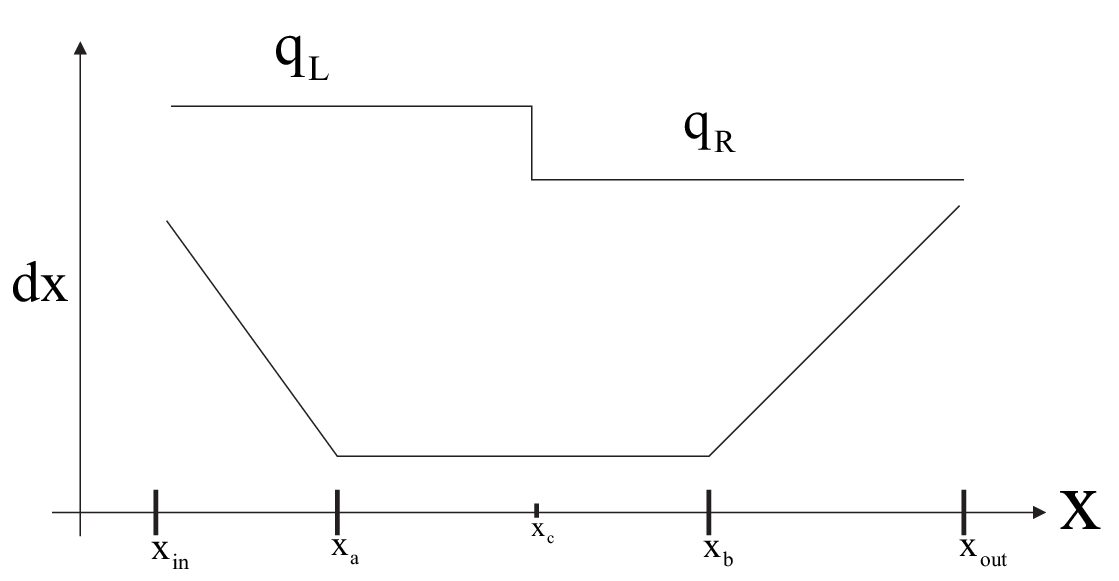}}
\caption{\small Grid spacing versus position. The interval of calculations is a priori divided into
two regions: $\rm [x_{a} \leq x \leq x_{b},$ where the physical variables are expected to change and
the sub-intervals  ${\rm x \leq x_{a}}$ and ${\rm x \geq x_{b}},$ where the physical variables do not change.
Therefore the grid spacing in the interval $\rm [x_{a} \leq x \leq x_{b}]$ is set to be uniform, whereas
in the rest intervals only several non-uniform grid-points have been used to save computer time.
  The initial conditions for the first test problem read: for $x\leq x_{c}:$  $\rm P_L = 40/3$, $\rm \rho_L = 10$ and
  for $\rm x > x_{c}:$ $\rm P_R = (2/3)\times 10^{-6}$ and $\rho_R = 1.$} \label{Fig2}
\end{figure*}


In Equation (\ref{Eq1}), the energy equation describes the time-evolution of the total
energy $\mathcal{E}^\mm{tot}$.
 Our numerical solution method relies first on selecting the primary variables and calculating
 their corresponding entries in the Jacobian. We then iterate
 to recover the correct contributions of the depending variables, such as the transport velocity and pressure. Therefore, the advection operator can be decomposed into two parts as follows:

 \beq
 \rm
{\D_t \mathcal{E}^\mm{tot} }+ \nabla_r \cdot (\mathcal{E}^\mm{tot} V^r)
                    = \nabla_r \cdot (P\, V^r).
                \label{Eq5T1}
\eeq
This equation is solved as follows: Solve for $\rm \mathcal{E}^\mm{tot}$ using the pressure from the
last iteration level. Once $\rm \mathcal{E}^\mm{tot}$ and ``D" are known, we may proceed
to update the pressure as follows:
\beq
\rm
 P = \DD{\mathcal{E}^\mm{tot} - u^t D}{\DD{\gamma}{\gamma -1}(u^t)^2 - 1}.
\eeq
This value of ``P" is then inserted again into the RHS of Equation (\ref{Eq5T1}) to compute a corrected value for
$\rm \mathcal{E}^\mm{tot}$.

The effect of iteration can be reduced on the cost of using pressure values from the last
time step using the following alternative formulation:

\beq
\rm
\D_t {\tilde{\mathcal{E}}^\mm{tot} }+ \nabla_r \cdot (\tilde{\mathcal{E}}^\mm{tot} V^i)
                    = {\D_t P },
                \label{Eq5T2}
\eeq
where $\rm \tilde{\mathcal{E}}^\mm{tot} = (u^t)^2 \rho h.$  \\
Thus, once $\rm \tilde{\mathcal{E}}^\mm{tot}$ is found, the pressure
can be computed from the known conservative variables as follows:
\beq
\rm
 P = \DD{\gamma - 1}{\gamma}\DD{\tilde{\mathcal{E}}^\mm{tot} - u^t D}{(u^t)^2}.
\eeq

On the other hand, using the internal energy formulation for modeling moving shocks requires
the inclusion of an artificial viscosity for calculating their fronts accurately.
This modification is necessary in order to recover the loss of heat generally produced through
the conversion of kinetic energy into internal energy at the shock fronts. \\
In this case,   the RHS of Equation (\ref{Eq5in}) should be modified to include
the artificial heating term:
\beq
\rm
Q^+_{art} = \eta_{art}({\D_\mu} u^t V^\mu)^2, ~~ \mu=0,1,2,3
\eeq
where $\eta_{art}$ is the artificial viscosity coefficient.\\
 The inclusion of  $Q^+_{art}$ implies an enhancement of the effective pressure. Therefore, the thermodynamical pressure,
 and  also the enthalpy, should be modified to include an artificial pressure of the form:
\beq
\rm
P_{tot} = P + P_{art} = P + \eta_{art}{\D_\mu}( u^t V^\mu).
\eeq
 Note that the artificial pressure enters the momentum equation in the form of $\nabla P_{art}$,
 which scales as $\sim \Delta (\eta_{art} V).$ This is equivalent to activating a second order
 viscous operator at the shock fronts, whose effect is then to transport information from the downstream
 to the upstream regions, so to enhance the stability of the transport scheme in such critical
 transitions.

\begin{figure*}[htb]
\centering {\hspace*{-0.2cm}
\includegraphics*[width=13.95cm, bb=-150 215 760 595,clip]
{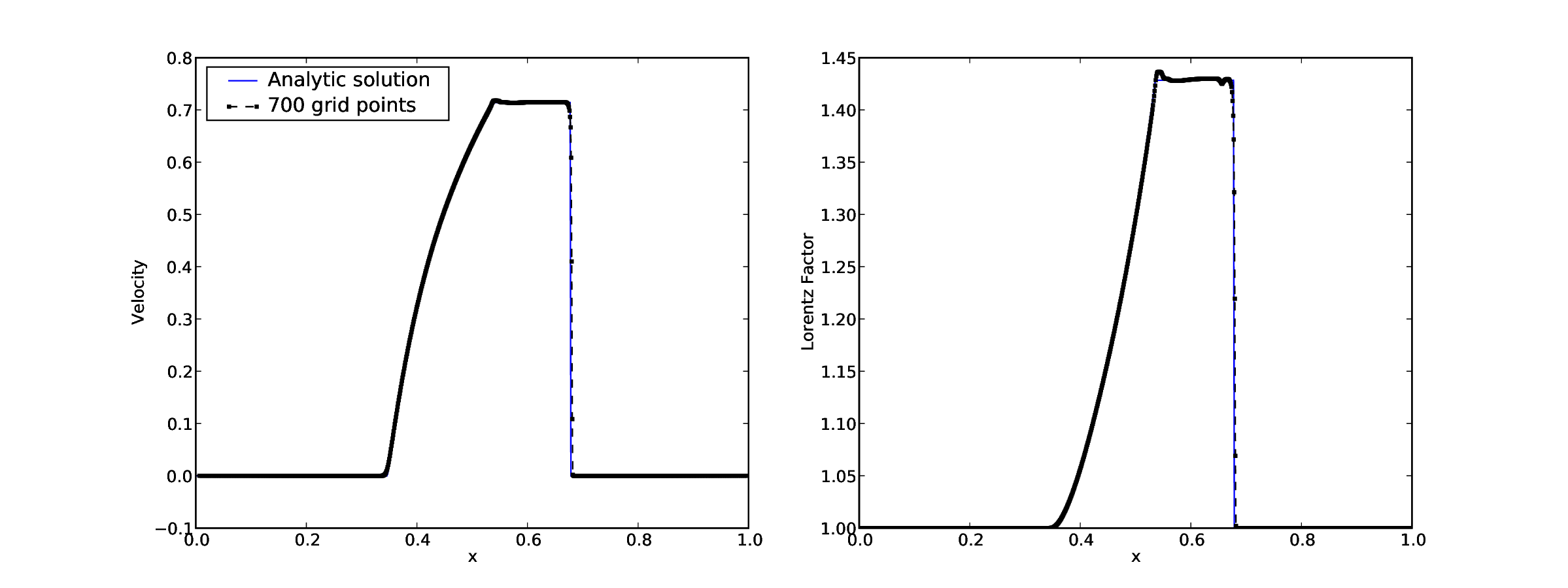}}
\caption{\label{Fig1} \small The profiles of the velocity $\rm V$ (left) and the Lorentz factor
         $\rm u^t$ (right, dashed line connecting squares) overplotted on the corresponding
          analytical solutions
          (continuous blue lines) after $\tau = 0.2$. The numerical solutions have been obtained
           using the total energy formulation (TEF). The initial pressure was taken to be
         $\rm P_{L}= 40/3,~\rho_L = 10,$ $\rm P_R= (2/3)\times 10^{-6},~\rho_R = 1.$ We use 700 non-uniformly distributed grid points, CFL=0.4 and perform two global iterations in each time step.
         The advection scheme employed here is of third order spatial and second order temporal accuracy.
} \label{Fig3}
\end{figure*}
\begin{figure*}[htb]
\centering {\hspace*{-0.2cm}
\includegraphics*[width=13.95cm, bb=-150 215 760 595,clip]
{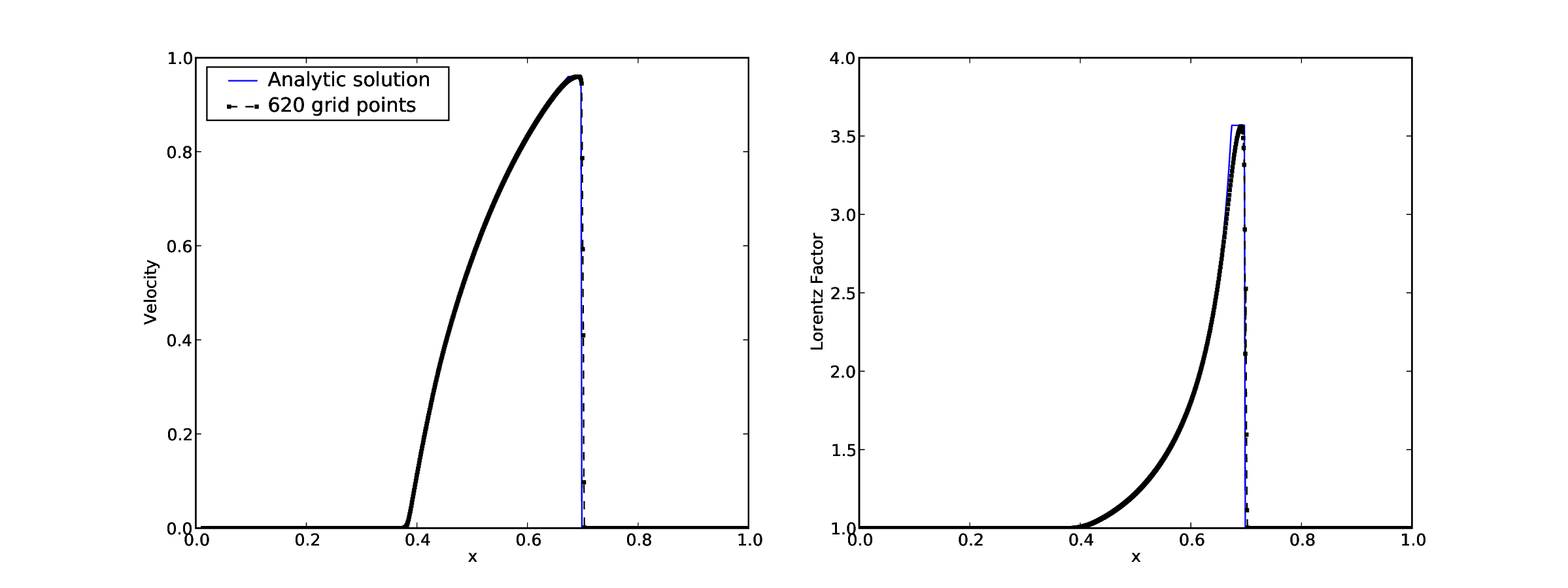}}
\caption{\small  As in Fig (3), the profiles of $\rm V$  and $\rm u^t$   have been obtained
           using the TEF, an initial pressure $\rm P_L= 10^2 P_{L0}$  and
           CFL= 0.4. 620 non-uniformly distributed grid points have been used { in combination with
           four global iterations per time step.}} \label{Fig4}
\end{figure*}

\subsection{Determining the Lorentz factor}
Our test calculations showed that the Lorentz factor can be best determined from the normalization
condition, in which both the conservative variables and transport velocities are involved. To clarify
this point, the normalization condition reads:
\beq
\rm
\barr{ll}
-1 & =  u^\mu u_\mu \\
   & =  (u^t V^\mu)~ (\DD{M_\mu}{\bar{D}})\\
   & =  \DD{u^t}{\bar{D}}[M_t + V^r M_r + V^\theta M_\theta + V^\varphi M_\varphi ] \\
   & =  \DD{u^t}{\bar{D}}[\{\DD{M^t - g^{t\varphi}M_\varphi}{g^{tt}}\} + V^\theta M_\theta + V^\varphi
          M_\varphi] \\
   & =  \DD{u^t}{\bar{D}}[ (\DD{\bar{D}u^t}{g^{tt}})  + V^\theta M_\theta
         + (V^\varphi - \DD{g^{t\varphi}}{g^{tt}})M_\varphi], \\
\earr
\label{Eqnorm}
\eeq
where the velocities are in light speed units, $\rm \bar{D} = \rho h u^t,$ which can be directly determined from the continuity and the energy
equations, $\rm V^\mu$ is the transport velocity and $\rm g^{\mu\nu}$ are the elements of the metric.

The final form of Eq. (\ref{Eqnorm}) is the quadratic equation: $\rm A~(u^t)^2 + B\,u^t +C = 0, $
where A, B and C are parameters independent of $u^t.$ Thus,
the Lorentz factors can be determined from this equation completely.\\
We note that since the state of the conservative variables $\rm D,~M_\mu, {\mathcal{E}}^d$ depends strongly
 on the transport velocity $\rm V^\mu,$ it is therefore necessary to consider $\rm V^\mu$ when
computing the Lorentz factor $u^t$ as described in Eq. (\ref{Eqnorm}).
\section{The numerical method: advection scheme for high Lorentz factors}

{The advection scheme presented in Sec. (3.1) is incorporated and solved using the implicit general relativistic numerical MHD solver, GR-I-RMHD.  This solver has been described in details in
Hujeirat et al. (2008). For completeness, we briefly mention that GR-I-RMHD is a 3D axi-symmetric, time-implicit solver, which relies on using the powerful preconditioned defect-correction iterative strategy in combination with the nonlinear Newtonian iterative method. By preconditioning we mean constructing a matrix that is similar to the original Jacobian obtained by differentiating the equations with respect to the main unknown variables.  \\
The advantages of implicit over explicit methods are prominent when quasi-stationary or time-independent
flow configurations are sought. However, for strongly time-dependent flow features, such as moving shock and
turbulence, explicit methods are much more efficient, provided that flow is almost ideal, i.e., non-reacting and non-dissipative. Towards enlarging the range of applications of GR-I-RMHD to include the regime of
high Lorentz factors, the advection scheme presented in Sec. (3.1) has been incorporated and
verification tests have been pereformed (see Sec. 4).    }

\begin{figure*}[htb]
\centering {\hspace*{-0.2cm}
\includegraphics*[width=13.95cm, bb=-150 215 760 595,clip]
{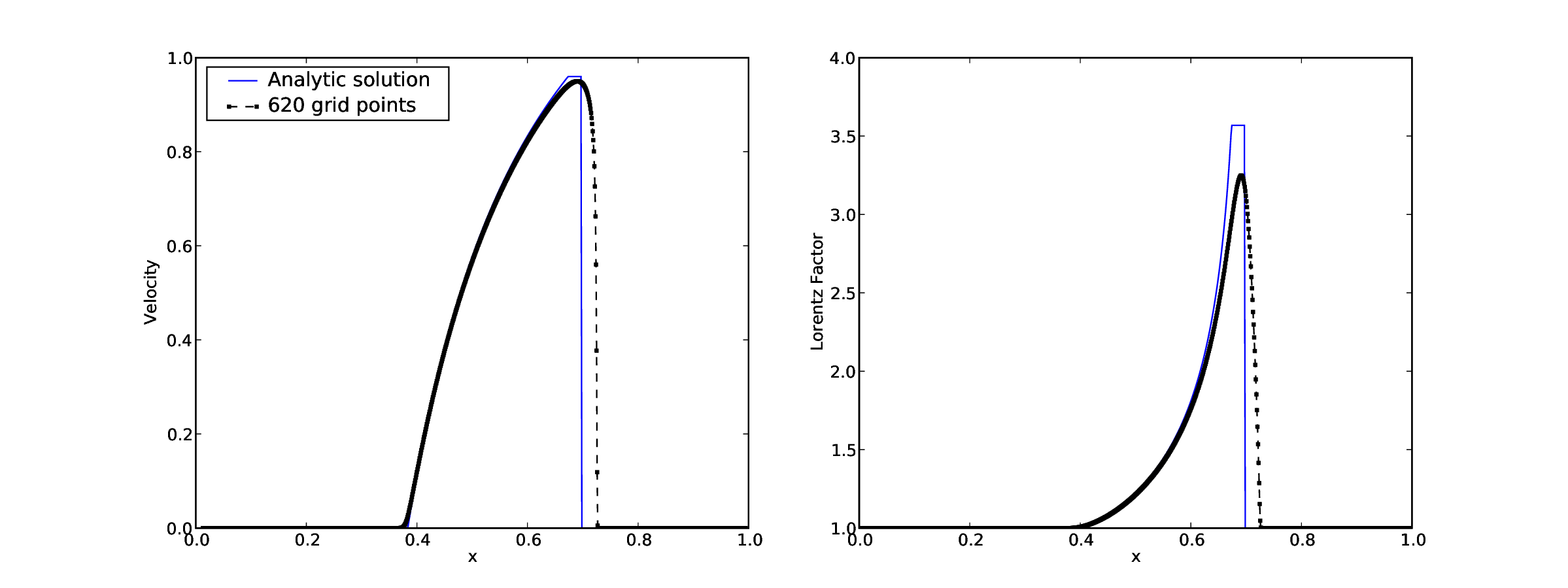}}
\caption{\small  As in Fig (3), the profiles of $\rm V$  and $\rm u^t$   have been obtained
           using the TEF, an initial pressure $\rm P_L= 10^2 P_{L0}$ and
            CFL= 0.4. An upwind advection scheme of first order spatial accuracy { and two global iterations
            per time step have been used.} } \label{Fig6}
\end{figure*}

\subsection{The third order advection scheme}
Most time-implicit advection schemes generally rely on upwinding to stabilize the transport near
critical fronts. Transport of quantities here can be mediated with CFL numbers larger than unity.
 Therefore, the advection scheme employed
should be independent of the time step size. Indeed, the monotone upstream centered schemes for
conservation laws, or simply the MUSCL schemes,
obey these conditions \citep[see][and the references therein]{Hirsch}.
However, in order to model moving shocks with high Lorentz factors accurately,
 it is necessary to modify the scheme, so that the information-flow from the downstream
to upstream regions should be
further limited in accordance with the normalization condition, i.e., causality condition.

A classical advection scheme of the MUSCL-type gives the following interface values:

\beq
\barr{ll}
 \rm q^{SR}_{j-1/2} = q_{j-1} - \DD{1}{4}[(1+\kappa) \Delta q_{j-1} + (1-\kappa) \Delta q_{j-2}]
 & \textrm{:~if~~~} V_j \leq 0 \\
 \rm  q^{SL}_{j-1/2} = q_{j}~ +~ \DD{1}{4}[(1-\kappa) \Delta q_{j} + (1+\kappa) \Delta q_{j+1}]
 & \textrm{:~if~~~} V_j > 0,
 \earr
 \label{Vanleer}
\eeq
where ``q" is the transported physical quantity, $\kappa $ is a switch off/on parameter that specify the accuracy needed and
$\rm \Delta q_{j} = q_{j} - q_{j+1}.$
 Note that the scheme is of second order for $\kappa = -1, 0, 1$ and of third order
  for $\kappa = 1/3$.

In order to enable an accurate capturing of shock fronts propagating with high Lorentz factors
on a non-uniform grid distribution, the following modifications have been performed:

 \beq
 \rm
\barr{l}
 \textrm{if~~~} V_j \leq 0:~~~
 \rm  q^{SR}_{j} = q_{j-1} - \DD{1}{4}[(1-\sigma) \Delta q_{j} + (1+\sigma) \Delta q_{j-1}],\\
 \textrm{~~~~~~where~~~} \rm  \sigma = 2 \kappa~(\Delta q^{RR}_{j}\cdot\Delta q^{0R}_{j})/
                            [(\Delta q^{RR}_{j})^2 + (\Delta q^{0R}_{j})^2 + \epsilon] .\\
 \textrm{if~~~} V_j > 0:~~~
\rm  q^{SL}_{j} = q_{j} + \DD{1}{4}[(1+\sigma) \Delta q_{j} + (1-\sigma) \Delta q_{j+1}],\\
\textrm{~~~~~~where~~~} \rm  \sigma = 2 \kappa~(\Delta q^{0L}_{j}\cdot\Delta q^{LL}_{j})/
                            [(\Delta q^{0L}_{j})^2 + (\Delta q^{LL}_{j})^2 + \epsilon], \\
\earr
\label{MVanleer}
\eeq
where
\beq
\rm
\barr{l}
\rm  \Delta q^{RR}_{j} = ({x^m_{j-1} - x^m_{j}})(q_{j-2} - q_{j-1})/({x^m_{j-2} - x^m_{j-1}})  \\
\rm  \Delta q^{0R}_{j} = ({x^m_{j-2} - x^m_{j-1}}) (q_{j-1} - q_{j})/({x^m_{j-1} - x^m_{j}})\\
\rm  \Delta q^{0L}_{j} = ({x^m_{j} - x^m_{j+1}}) (q_{j-1} - q_{j})/({x^m_{j-1} - x^m_{j}})  \\
\rm  \Delta q^{LL}_{j} = ({x^m_{j-1} - x^m_{j}}) (q_{j} - q_{j+1})/({x^m_{j} - x^m_{j+1}}),  \\
\earr
\eeq
and where $\epsilon$ is a small number set to avoid division by zero.


The accuracy of above modified scheme may be further increased
by incorporating the Lagrangian Upwind Interpolation Scheme (LUIS).
 The LUIS strategy is based on by constructing a Lagrangian polynomial of third or
fourth order whose main weight is shifted to the right or left, depending on the
upwind direction.
The final combined LUIS-MUSCL scheme reads as follows:
\beq
\rm
\barr{l}
 \textrm{if~~~} V_j \leq 0:~~~
 \rm  q^{SR}_{j} = \xi q^{SR}_{j} + (1 - \xi)Lq^{SR}_{j},\\
 \textrm{if~~~} V_j > 0:~~~
 \rm  q^{SL}_{j} = \xi q^{SL}_{j} + (1 - \xi)Lq^{SL}_{j},
\earr
\label{MMVanleer}
\eeq
where $\xi$ is an additional weighting function and $\rm Lq^{SR,SL}_{j} $ are LUIS corrections
of ``q" at the interface $\rm r_j$ (see Fig. \ref{Fig2}, which are computed in the following manner:

\beq
\rm
\barr{ll}
\rm  Lq^{SR}_{j} = \sum_{l=j-2}^{l=j}q_l \prod_{i\neq k,k=j-2}^{k=j} \DD{r_j - r^m_k}{r^m_i - r^m_k},
 &  \textrm{:~if~~~} V_j \leq 0 \\
\rm  Lq^{SL}_{j} = \sum_{l=j}^{l=j+2}q_l \prod_{i\neq k,k=j}^{k=j+2} \DD{r_j - r^m_k}{r^m_i - r^m_k},
&  \textrm{:~if~~~} V_j > 0.\\
\earr
\eeq

However, in order to maintain monotonicity of schemes,  high order advection schemes
 must degenerate into lower ones close to shock fronts,  which in turn enhance the production of numerical entropy/diffusion. Therefore, in most of the test calculations presented here it was necessary to
have $\xi>1/2$  in order to damp over- undershootings and enable convergence.
\begin{figure*}[htb]
\centering {\hspace*{-0.2cm}
\includegraphics*[width=13.95cm, bb=-150 215 760 595,clip]
{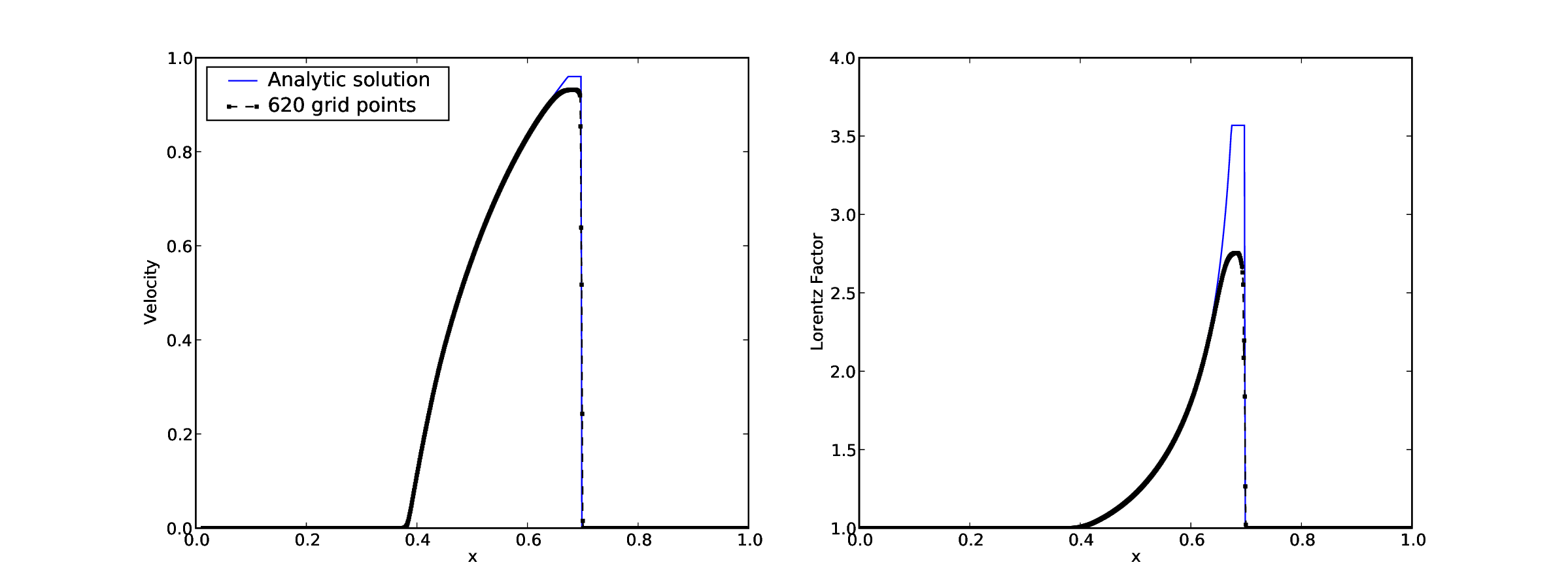}}
\caption{\small  As in Fig (3), the profiles of $\rm V$  and $\rm u^t$  here have been obtained
           using the TEF method together with a heating source due to the artificial viscosity.
            An initial pressure $\rm P_L= 10^2 P_{L0},$
           CFL= 0.4 and a spatially third order accurate advection scheme
           {in combination with two global iterations per time step have been used} } \label{Fig7}
\end{figure*}
\begin{figure*}[htb]
\centering {\hspace*{-0.2cm}
\includegraphics*[width=13.95cm, bb=-150 215 760 595,clip]
{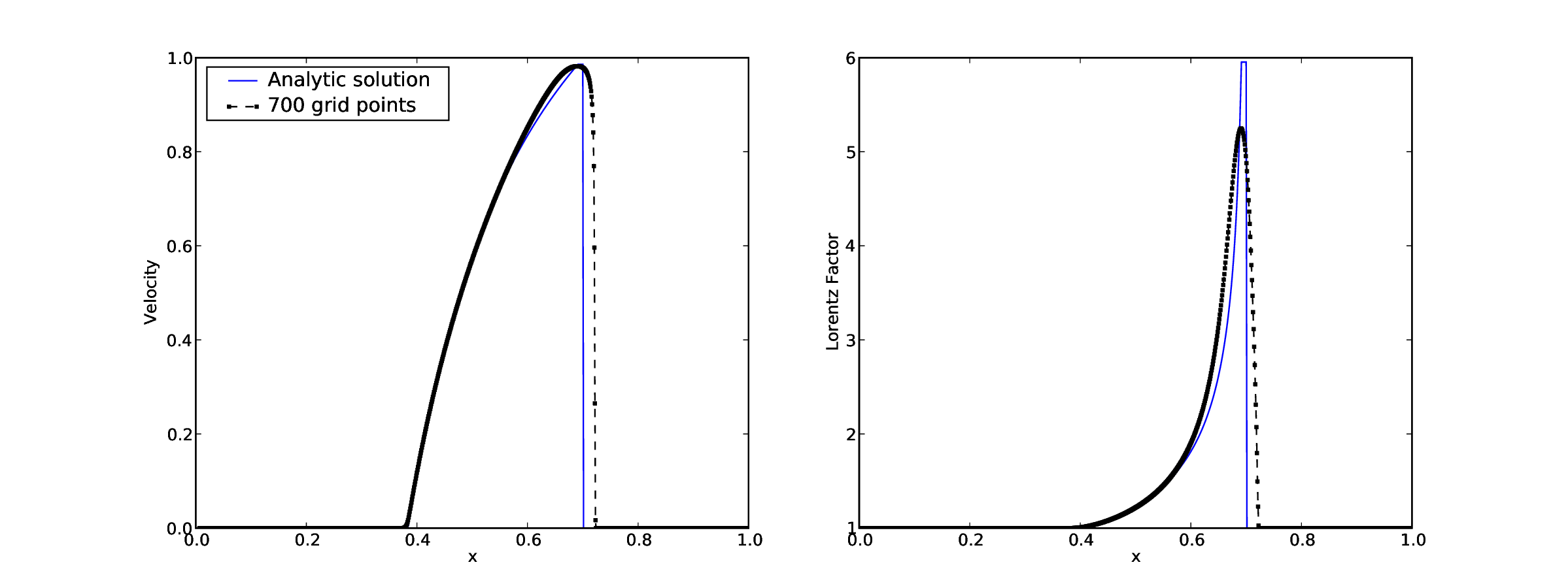}}
\caption{\small  As in Fig (3), the profiles of $\rm V$  and $\rm u^t$   have been obtained
           using the pseudo-total energy formulation (PTEF, see Eq. \ref{Eq5T2}),
           an initial pressure $\rm P_L= 10^3 P_{L0}$,
           CFL= 0.4 and four global iterations per time step.} \label{Fig8}
\end{figure*}

\section{Verification tests}

In this section we present results of various verification tests of the advection scheme using
different energy formulations. Although the scheme can be applied to fluid-transport in multi-dimensions,
the focus of the present study will be just to examine transport in one-dimension.
Thus, the equations to be solved in this case are the continuity, radial momentum, internal and the  total energy equations described in Eqs. (1)-(4).

The widely used hydrodynamical test in 1D  is the one-dimensional relativistic shock tube problem (henceforth
RSTP).
The advantages of this test problem is that the obtained numerical solutions can be
compared with the corresponding analytical solution with arbitrary large Lorentz factors.

For further details on the background of this test problem and the way the analytical solutions
are obtained we refer the reader to \citep{MartiMueller2003}.\\
The RSTP is an initial value problem, in which the solution depends mainly on the
initial conditions. The domain of calculations is divided into a right and left regions
(see Fig. \ref{Fig2}).
In the left region $\rm 0\leq x \leq x_{c}:$ we set $\rm P_L = 40/3$, $\rm \rho_L = 10$  and in the
right region $\rm x_{c}< x :$ we set $\rm P_R = (2/3)\times 10^{-6}$, $\rm \rho_R = 1.$
The initial velocity is set to vanish everywhere.
\begin{figure*}[htb]
\centering {\hspace*{-0.2cm}
\includegraphics*[width=13.95cm, bb=-150 215 760 595,clip]
{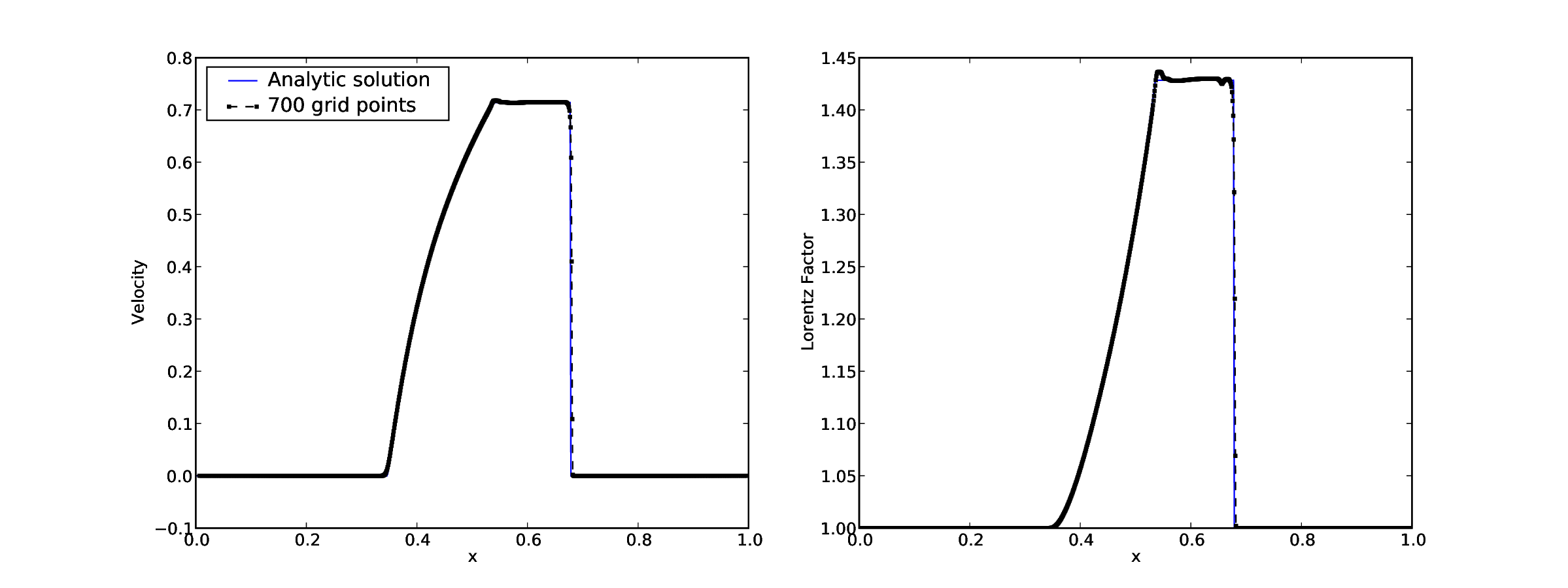}
 }
\caption{\small  As in Fig (3): the profiles of $\rm V$  and $\rm u^t$   have been obtained
           using the internal energy formulation (IEF, see Eq. \ref{Eq5in}),
           the initial pressure $\rm P_L= P_{L0},$
           CFL= 0.4  {and in combination with two global iterations per time step}. } \label{Fig9}
\end{figure*}
\begin{figure*}[htb]
\centering {\hspace*{-0.2cm}
\includegraphics*[width=13.95cm, bb=-150 215 760 595,clip]
{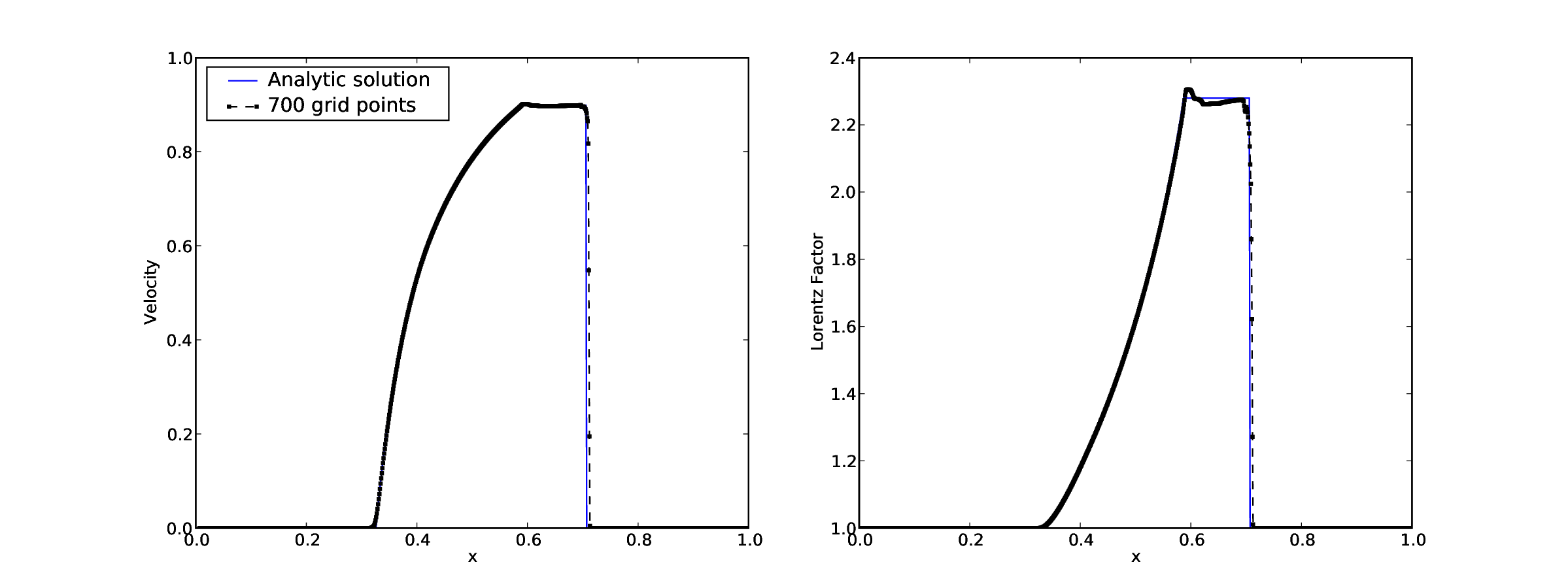}}
\caption{\small  Similar to Fig (3), the profiles of $\rm V$  and $\rm u^t$   have been obtained
           using the IEF with the initial pressure
            $\rm P_L= 10 P_{L0}$ {in combination with two global iterations per time step}. The rest of variables remained as in Fig. (1) unchanged.} \label{Fig10}
\end{figure*}

Using the total energy formulation  (TEF) we show in Figures (\ref{Fig3}-\ref{Fig13}) the profiles of the
velocity and the corresponding Lorentz factor for different distributions of the initial
conditions.\\
In Fig.\ref{Fig1} we show the profiles of the velocity and Lorentz factor after 0.2 sec
using the total energy formulation (see Eq. \ref{Eq1}). {To reduce computer time, the interval $\rm [0.35 \leq x \leq 0.75]$
has be divided into 400 equally spaced cells, whereas 220 cells were used to
cover the external intervals, where the variables continue to acquire their
initial values. This grid distribution is set initially and remains fixed in time.}
\begin{figure*}[htb]
\centering {\hspace*{-0.2cm}
\includegraphics*[width=13.95cm, bb=-150 215 760 595,clip]
{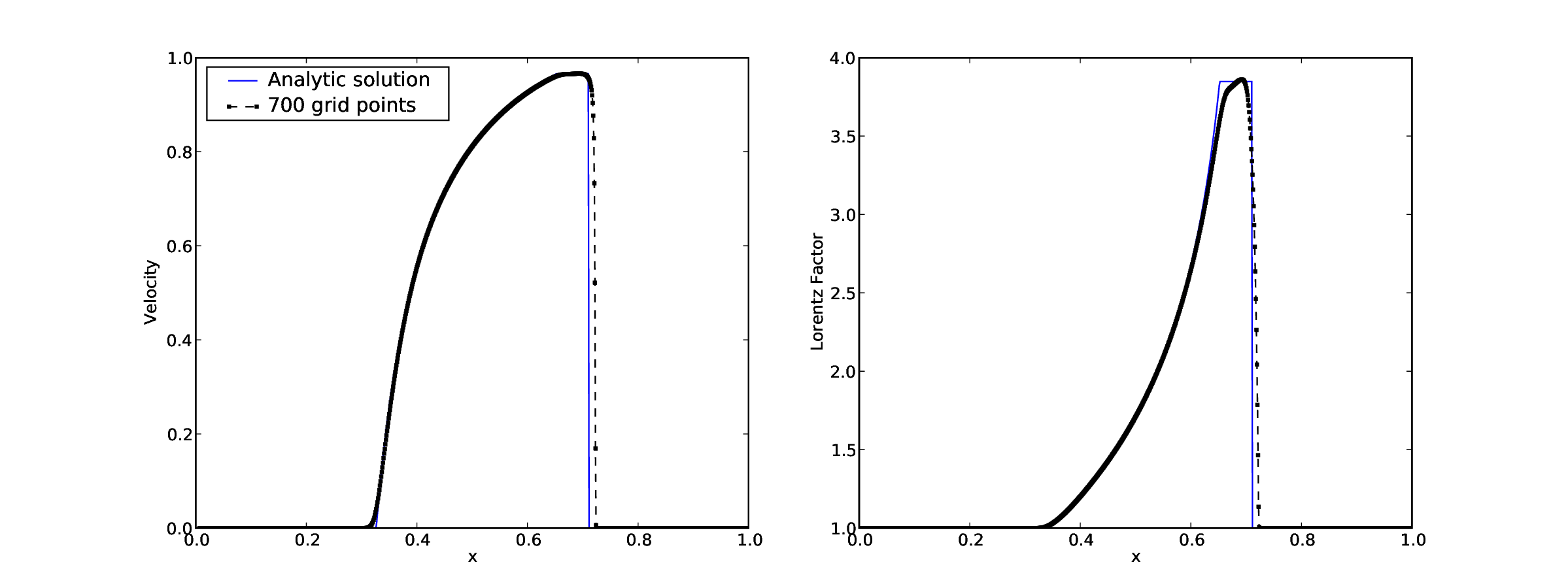}}
\caption{\small  As in Fig (3), the profiles of $\rm V$  and $\rm u^t$   have been obtained after $\tau = 0.215,$
           using the IEF, an initial pressure  $\rm P_L= 10^2 P_{L0},$
            CFL = 1.5  {and four global iteration per time step} } \label{Fig12}
\end{figure*}

 The advection scheme  described in (Eq. \ref{MMVanleer}) has been employed
to achieve $\rm 3^\mathrm{rd}$ order spatial accuracy and the
damped Crank-Nicolson method is used to achieve second order temporal accuracy \citep{HujeiratRannacher2001}.
The time step size is set to have a maximum that corresponds to CFL= 0.4.

To test the robustness of the solution procedure, we have increased the $\rm P_L$ by a factor
of 100, while the other variables remained unchanged.
This increase of the pressure yields the Lorentz factor: $\rm u^t \approx 3.6 $ (Fig. \ref{Fig4}).
However,  to enforce convergence, it was necessary to double the number of iteration
per time step considerably. \\
{ Comparing these results with those obtained using a double number of grid points and
a halved time step-size, we could not observe a visibly significant improvement of the results.}

From Figs. (\ref{Fig6}) and (\ref{Fig7}), we see that incorporating artificial viscosity
and artificial heating while using the total energy formulation (TEF) has a similar effect as that of the normal
numerical diffusion.
In both cases the velocity at the shock front attains lower values than expected from the analytical
solution.
\begin{figure*}[htb]
\centering {\hspace*{-0.2cm}
\includegraphics*[width=13.95cm, bb=-150 215 760 595,clip]
{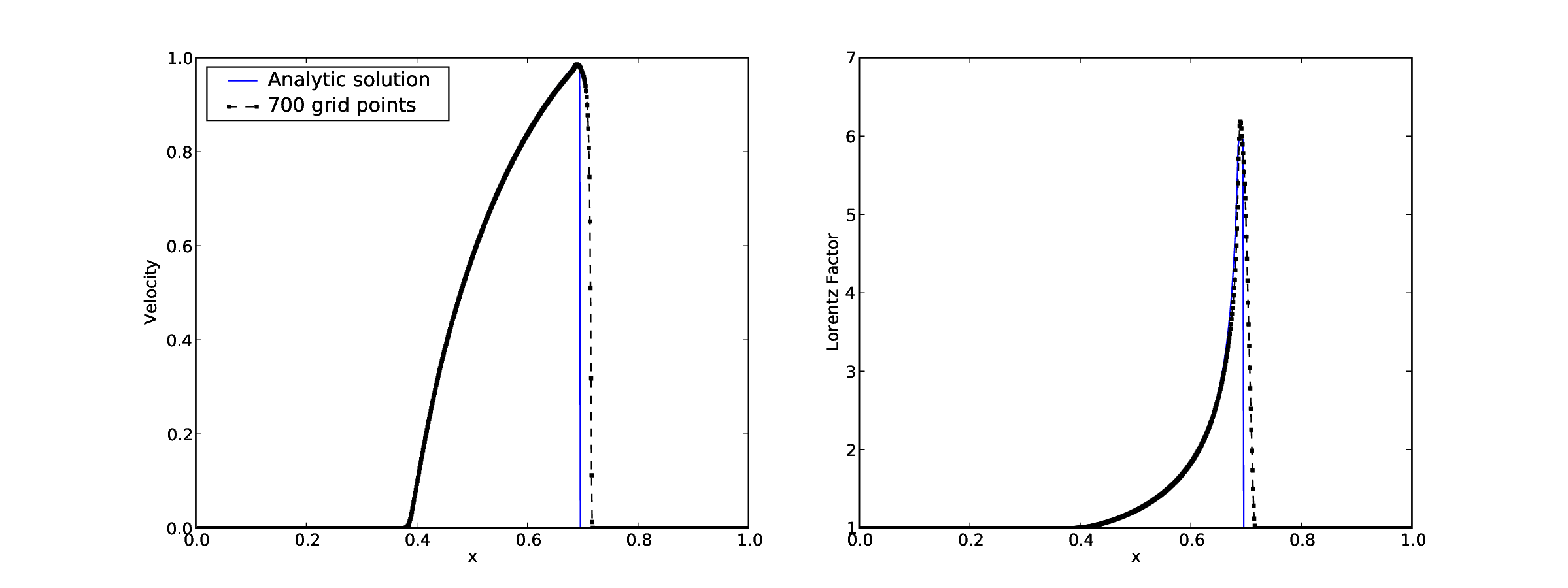}}
\caption{\small  As in Fig (3), the profiles of $\rm V$  and $\rm u^t$   have been obtained
           using the IEF method, an initial pressure  $\rm P_L= 10^3 P_{L0},$
            CFL= 2 in combination with five global iterations per time step. } \label{Fig12}
\end{figure*}
\begin{figure*}[htb]
\centering {\hspace*{-0.2cm}
\includegraphics*[width=13.95cm, bb=-150 215 760 595,clip]
{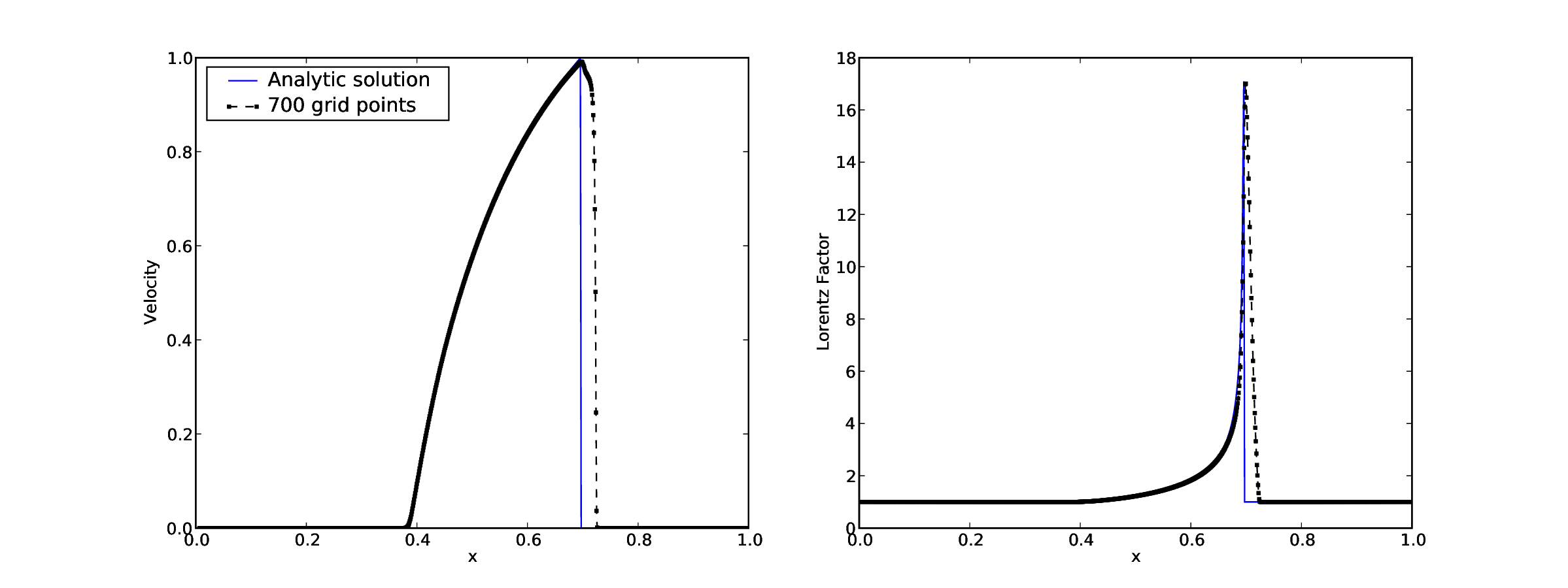}}
\caption{\small  As in Fig (3), the profiles of $\rm V$  and $\rm u^t$   have been obtained
           using the IEF method, an initial pressure $\rm P_L= 10^5 P_{L0},$
            CFL= 4 in combination with seven global iterations per time step. } \label{Fig13}
\end{figure*}
\begin{figure*}[htb]
\centering {\hspace*{-0.2cm}
\includegraphics*[width=13.95cm, bb=-150 215 770 595,clip]
{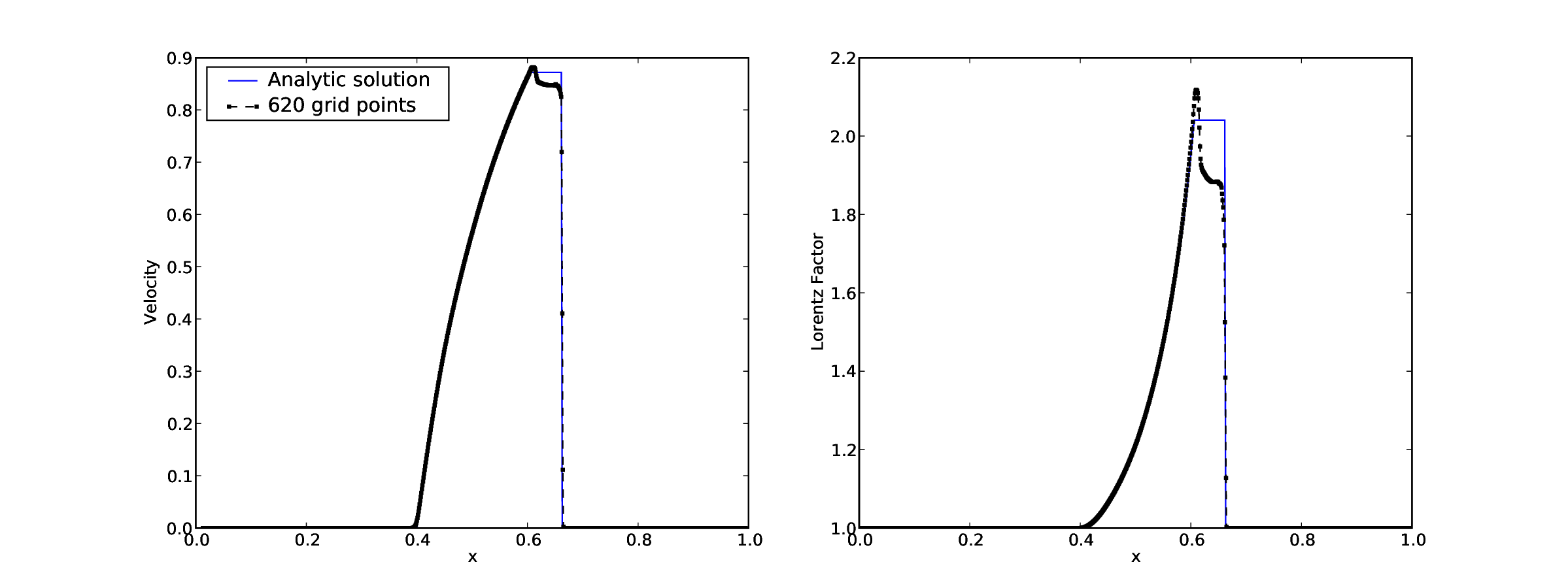}}
\caption{\small  As in Fig (3), the profiles of $\rm V$  and $\rm u^t$  are shown after $\tau = 0.175.$ These have been obtained
           using the compact internal energy formulation (CIEF, see Eq. \ref{Eq5in2}),
           initial pressure $\rm P_L= 10 P_{L0}$  {in combination with two global iterations per time step} } \label{Fig14}
\end{figure*}

We note that within the context of time-implicit solution methods,
the total energy approach has a limited capability to accurately capture moving
shocks at high Lorentz factors (Fig. \ref{Fig8}). The reason is that, unlike the
internal energy $\rm E_{in}$,  the kinetic energy $\rm E_{kin}$ increases with Lorentz factor as $\Gamma^2.$
As a consequence, the convergence rate decreases with $\rm E_{in}/E_{kin}.$ In this case
it is necessary to decrease both the time step size and the grid spacing  in addition to allow an
increase of the number of iteration per time step. The actual number of iteration
per time step required to assure convergence is residual-dependent and in most cases cannot be determined a priori. Our test calculations with large gamma-factors  showed a sharp increase of the actual number of iteration per time step (e.g., from 4 iteration for $\Gamma \approx 3.6$ to 12 for
$\Gamma \approx 6$) or even a stagnation of the solution method.

On the other hand, our time-implicit solution method is found to be more robust if the
internal energy formulation (IEF, see Fig. 2) is used. In Figures (\ref{Fig9}-\ref{Fig13}), the results
of several test calculations are shown, which agree well with the analytical solutions.
Moreover, the IEF method is numerically stable and the implicit solution procedure converges
even when running the calculations with  $\rm CFL\geq 1$ (see Figs. \ref{Fig12} and \ref{Fig13}).

The major draw back of the IEF is the necessity to fine-tune the coefficients of the
artificial viscosity in order to obtain the correct Lorentz factor that fits well with the analytical solution. By fine-tuning we mean a posteriori adjustment of the numerical  to the analytical solution
by carefully fitting the artificial viscosity coefficient so to attain the best agreement between both.  \\
Moreover, the fine-tuning procedures must be repeated each time the
ratio of the pressure $\rm P_L/P_R$ is changed. This is a consequence of the conversion of kinetic energy
into heat at the shock fronts which cannot be determined a priori and uniquely by solely solving
the internal energy equation.

Nevertheless, using the IEF, non-linear physical processes can be easily incorporated into
the implicit solution procedure. Unlike in the TEF case, such processes are
direct functions of the internal energy and therefore it is much easier to compute their
entries and incorporate them into the corresponding Jacobian.
 This has the consequence that calculations can be performed also with $\rm CFL \geq 1 $,
 while still maintaining  consistency with the original physical problem.

Finally, in Fig. (\ref{Fig14}) we show the profiles of $\rm V$ and $\rm u^t$ using
 the compact internal energy formulation (CIEF, Eq.(\ref{Eq5in2}) with high accuracies.
The method is found to display overshooting and undershooting both at the contact discontinuity and
the shock front, whose appearing was difficult to prevent through enlarging the number of grid points, decreasing the time step, enhancing the number of iteration per time step or even by activating the artificial viscosity.\\
It should be noted here that since in the limit of large Lorentz factors, the CFL number depends
weakly on the sound speed (see Fig. \ref{Fig15}), increasing {the number of global iterations} per
time step has more significant effect on convergence than merely decreasing the time
step size.\\

{To examine the stability of the advection scheme at extremely high Lorentz factors, i.e.,
$\Gamma \geq 10$, we applied the solver to the relativistic planer shock reflection
problem (RPSR), which has been comprehensively discussed by \citet{Aloy1999}.  \\
The RPSR problem is based on simulating the collision of two relativistically, oppositely directed  and
equally fast moving cold plasmas. After they collide, an intermediate region of motionless, hot and dense  matter
is formed at the center.\\
The following initial conditions are used:
$\rho_L = \rho_R = 1,~~ \epsilon_L = \epsilon_R = 2.29\times 10^{-5}$ and allow the plasmas to
collide with $V_L=-V_R$ at the middle point, i.e., at x=0.5 in Fig (\ref{Fig15}).
We then gradually increase $V_R$ to reach almost the speed of light.\\
In Fig (\ref{Fig15}), the profiles of the density of the shocked motionless matter in the intermediate region, that have been
obtained using different collisional speeds, are shown.
As expected, the post-shock density in the intermediate region correlates with the collisional speed of the plasma.
Although our results agree well with those reported by \citet{Aloy1999} for $V<0.999999$, our advection scheme
seems to diverge for $V>0.99999.$  It shows some sort
 of sub-grid oscillations associated with under and overshootings both at the center as well as
 at the front of the reflected shock.
 We attribute this problem to the enhanced numerical errors in the evaluation of the
 Lorentz factor  from the normalization condition as well as from the inverse non-linear transformation
 between the conservative and primitive variables.  Our attempts in this regime to further optimize the shock capturing
 scheme or using other formulations of the normalization condition produced non-noticeable improvement to
 the results.\\

   Nevertheless, the test calculations show that our numerical procedure presented here is
 robust and capable of dealing accurately with ultra-relativistic velocities that
 govern the fluid-motions in diverse astrophysical events, such as in quasars or jetted Gamma-ray bursts.
  However, testing the advection scheme for velocities corresponding to $\Gamma > 1000$ is physically irrelevant, as the fluid-description of plasmas breaks down.}
\begin{figure}
\centering {\hspace*{-0.2cm}
\includegraphics*[width=8.5cm, bb=92 69 577 454,clip]{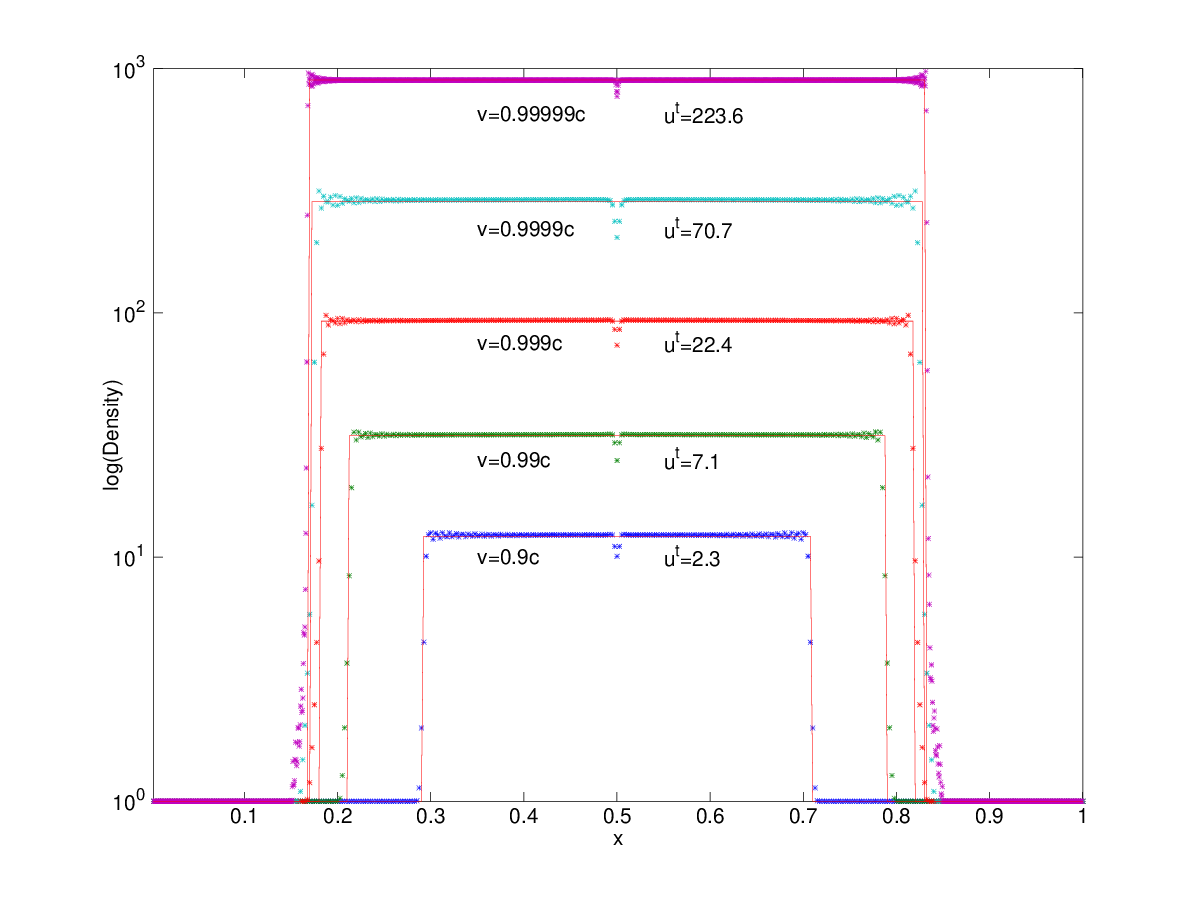}}
\caption{\small   Shock reflection problem: various density profiles  resulting from  different
collisional speeds (v in units of the speed of light) versus the distance from the collisional center x=0,
using 400 grid points  and  CFL=0.3 after the elapsed time $t=1$. The corresponding Lorentz factor
  $u^t$ is assigned to each profile.  } \label{Fig15}
\end{figure}
\section{Summary and conclusions}

In this paper we have presented a numerical scheme for time-implicit modeling of
relativistic shocks with high Lorentz factors. The results of several verification
tests using the internal energy (IEF), the total energy (TEF) and  pseudo-total
energy (PTEF) formulations have been presented as well.

To be stressed here that time-explicit methods are efficient and best suited for modeling strongly time-dependent
flow-problems much more than their implicit counterparts.
This is to be attributed to the fact that in implicit methods the coefficients of
preconditioning matrix should be evaluated and inverted several times per time step,
in order to recover the non-linearities accurately.
Furthermore, when modeling moving shocks or strongly time-dependent flows, a
time step-size which corresponds to $CFL < 1$ should be used to assure accurate capturing of the
temporal behaviour of the flows in critical regions.

On the other hand, implicit solvers are advantageous  over explicit solution methods when dealing
with reacting flows with physical processes operating on shorter or longer time scales compared
to the dynamical one or when the flows are viscous-dissipative.    Therefore, our attempt in this paper is
to enlarge the range of applications of the solver GR-I-RMHD to cover the regime of moving shocks at relatively high Lorentz factors.

Our test calculations show that TEF is most suited for modeling the propagation
of shocks with low to moderate Lorentz factors, though with  low CFL numbers,
hence more suitable for time-explicit solution methods.

On the other hand, although the PTEF is similar to the TEF, it was found that
a larger number of iterations per time step was required to maintain convergence.

Using the IEF however, it has been verified that the implicit solution procedure
converges relatively fast even for Lorentz factors $~\rm u^t \gg 1 $  and the $\rm CFL\geq 1$.\\

It should be stressed here that the IEF has a major draw back, as it requires a fine-tuned
parameters of the  artificial viscosity to obtain correct Lorentz factors at the shock
fronts. On the other hand, when modeling astrophysical plasmas, the IEF is favorable
over TEF, as cooling and heating processes are generally direct functions of the
internal energy, and so it is easier to linearize and construct the coefficients to be
 incorporated into the Jacobian.\\

{
Furthermore,  turbulence is an intrinsic property of most astrophysical
fluid flows. The artificial viscosity in such flows is redundant when turbulent diffusion
is taken into account, therefore relaxing  the fine-tuning problem of the artificial viscosity.
In the case of  ideal flows (; non-viscous, magnetic and radiative non-diffusive flows),
the solution procedure should run similar to the those adopted in the present test
calculations together with a re-start option in combination with a careful data-storage strategy.
Having started the calculations with an enhanced artificial viscosity, the calculation can then be
re-started after a certain elapsed time while  gradually decreasing the viscosity coefficient till
the optimal value is reached.  \\
Nonetheless a fine-tuned artificial viscosity may turn out to be  necessary also for numerical schemes using the
total energy formulation, in particular when shock-formation in the vicinity of relativistic objects
is to be modeled. In this case the ratio
of the internal to the total energy can be as low as $\delta t$ or even smaller. Using  highly accurate
non-diffuse numerical advection schemes may easily lead to over/under-shootings of the values
of the conservative variables at the shock
front\footnote{ It should be noted that most highly accurate advection schemes
generally violate the monotonicity condition in such critical regions \citep{LeVeque1990}.}
that give rise to negative temperatures, hence to a stagnation of the solution procedure. \\

We note that the phenomenon of shock propagation with high Lorentz factors has been poorly studied numerically,
 despite the availability of relatively large number of relativistic explicit codes. Most of these numerical
studies rely on using the TEF-method and the results obtained are indeed highly accurate
\citep[see for example][]{Hawleyetal1984, Aloy1999, DeVilliersHawley2003, Gammie_etal2003, Anninos, DeColle2011}.
On the other hand, to our knowledge PLUTO is the only solver that has been verified to accurately capture shock fronts propagating
with high Lorentz factors, i.e., $\Gamma > 5$ \citep{Mignone_etal2007}. For $\rm P_L = 10^3 P_R,$ PLUTO results
agree well with our results displayed in Fig.\ref{Fig12}.
For much higher $\Gamma-$factors we strongly recommend using an implicit method in combination with the internal energy
formulation.

While simulating the propagation of ultra-relativistic shock-fronts in 1D using implicit methods
is to date a feasible task on personal computers, the situation in 2D may still be different.
For example, assume we want to estimate the computational costs associated with modeling the
motion of 2D-curved relativistic shocks in the vicinity of a black hole (e.g.,  with $\Gamma = 17$ as in Fig.\ref{Fig13}).
The computational domain should consist  of approximately $N\times N =1000\times 1000$ non-uniformly distributed grid points.
The corresponding pre-conditioner would have a band width: $m_{bw} \approx 15. $ The computational costs per iteration
 would roughly be of the order $CC \approx m_{bw}\times N^2$. Taking into account that approximately
10 iterations per time step are needed and that a total number of about $10^5$ time steps are required in order to recover several dynamical time scales, we end up with approximately $10^{13}$ arithmetic operations. This number, however,
may be modified considerably when the overhead operations are taken into account. It may even increase by additional
 seven orders of magnitude, if the third spatial dimension is included or when magnetic fields and/or radiation are
 taken into account.\\
Therefore,  carrying out such calculations is indeed a challenging endeavor. \\

}
Concerning the evaluation of the Lorentz factor, we find that irrespective of the method used for solving the energy equation, the Lorentz factor is
found to be best computed from the normalization condition in which the transport velocities
and momenta are used.\\

We have also presented a modified MUSCL advection scheme of third order spatial accuracy,
which has been constructed to enable accurate capturing of relativistically moving shocks.
The  accuracy of the scheme can be further enhanced by incorporating
the Lagrangian Upwind Interpolation Scheme (LUIS). \\
{Nevertheless, for saving computer time, we recommend using the LUIS
 scheme when strongly time-dependent flows with prominent shock fronts are to be simulated.  } \\

Noting that the
appropriate parameters of the artificial viscosity cannot be
determined a priori, obtaining the correct Lorentz factor may turn out to be a
difficult task.
In the case of standing shocks, the hierarchical solution scenario in combination with the defect-correction
iteration procedure \citep{Hujeirat2005} can be employed to gradual-enhance the accuracies of the scheme and obtain
the correct Lorentz factors at the shock fronts.

In a forthcoming paper, we intend to study the formation of strong
shocks in  curved spacetime near the surface of ultra-compact neutron stars.\\

{\bf Acknowledgment}
      SF thanks the SNF for the financial support.


\end{document}